# Next Generation Radio Astronomy Receiver Systems

## Astro2010 Technology Development White Paper

**Principle Author: Matthew A. Morgan**
National Radio Astronomy Observatory
434-296-0217
matt.morgan@nrao.edu

**Co-author: J. Richard Fisher, NRAO**

## Abstract

Ground-breaking radio astronomy observations in the coming decade will require unprecedented levels of sensitivity while mapping large regions of space with much greater efficiency than is achieved with current telescopes. This requires new instrumentation with the greatest achievable sensitivity, dynamic range, and field of view. Receiver noise is quickly approaching fundamental limits at most radio wavelengths, so significant gains in sensitivity can only be made by increasing collecting area. Mapping efficiency requires radio telescopes with wider fields of view. Jointly, these requirements suggest using large arrays of smaller antennas, or many moderate-size antennas equipped with multi-beam arrays.

The engineering community is thus challenged to develop receivers and wide bandwidth data transport systems which are lower cost, more compact, more reliable, lower weight, more reproducible, and more stable than the best current systems, *with no compromise to performance*.

This can be achieved with a greater degree of component integration, extensive use of digital signal processing and transport, and replacement of functions currently performed in expensive and bulky waveguide and coaxial cable components with digital arithmetic and thin optical fibers. There are no miracles to be pulled from the technological hat. All of this is to be performed with careful attention to detail and adoption of the latest products from the consumer and industrial electronics industry.

In this white paper, we outline the complete redesign and re-optimization of receiver architecture to take advantage of the latest advancements in commercial technology. As well as transferring certain critical functions from the analog domain into the digital domain, this inevitably involves the seamless integration of the conversions from RF to baseband, from analog to digital, and from copper to fiber within a single receiver module. The result is a well-optimized modern receiver architecture that is compact, inexpensive, reliable, and mass producible without compromising performance or versatility in any way.

## Science Drivers

Science enabled by greater sensitivity and field of view includes pulsar surveys of the entire volume of our own Galaxy, wide-field maps of dust and many molecular species in star-forming regions, searches for transient sources on time scales from milliseconds to years, and a full-sky survey of emission from neutral hydrogen in galaxies out to redshifts approaching z = 1 to determine the fundamental structure and evolution of the universe.

Radio astronomy has made enormous strides in sensitivity and observing efficiency in its seventy year history, but we still observe only a tiny fraction of the sky at a limited range of frequencies at any one time. As collecting areas grow and sensitivities increase with better receivers and greater signal processing bandwidths the radio sky becomes less empty. Results from recent all-sky surveys at Parkes, Arecibo, and the VLA have shown that not only do they enhance the statistics on known objects, but new types of objects, such as pulsar-pulsar binary stars and extremely powerful transient sources in the distant universe, wait to be discovered. Searches for and discovery of new millisecond pulsars will provide more probes of the nanohertz spectrum of gravitational waves.

Multi-beam and wide-field radio telescopes will enhance the accuracy of very long baseline interferometric astrometry by providing simultaneous measurements of target objects and nearby phase calibrators. VLBI astrometry is pushing trigonometric parallax measurements into the depths of our Galaxy and eventually beyond, observing classes of objects that are invisible to optical astrometry instruments.

Achieving greater sensitivity and observing efficiency with more collecting area and greater fields of view is not enough. To access the science, radio astronomers of the next decade must have even tighter control of the weak systematic errors of their instruments and the ability to accurately correct for them. The calibration and stability of the system is therefore critical.

Among the most challenging heterodyne receiver applications are large format focal plane arrays where it is desirable to pack the receiving antenna elements as close together as is physically and electromagnetically possible. For example, a feed-horn array for 100 GHz ($\lambda$ = 3 mm) on the Green Bank Telescope (GBT) with secondary focus F/D = 2 would have the horns spaced about 2.3 cm center-to-center in a hexagonal pattern. Each receiver element would need to fit in a 2.3-cm diameter cylinder behind the horn. In a prime focus "beam-forming" (phased) array at 10 GHz ($\lambda$ = 3 cm) the array elements would be approximately 2 cm apart. Both of these applications are beyond the current state of the receiver art, but they are well defined targets for research and development in the coming decade.

## Complete Redesign, One Function at a Time

In this white paper we define a radio astronomy receiver system to include all components from the point where an electromagnetic signal is confined to a waveguide or transmission line to the point where it is delivered to the input of a signal processor for correlation with other signals or for the extraction of intensity information as a function of time and frequency.

The stability and accuracy of digital signal representation has been recognized since the demise of the analog computer more than 50 years ago. Radio astronomy has been slow to take full advantage of this accuracy because its information bandwidths have generally been wider than the computation rate of digital hardware. Extreme approximations, such as one- or two-bit correlators have been feasible for about 40 years because of the weak noise-like character of celestial signals, but the full potential of multi-bit signal processing at bandwidths greater than a few tens of megahertz has been affordable only in the last ten or fifteen years. Analog to digital converters (ADCs) with 8-bit (256-level) signal representation are now available up to at least 10 giga-samples per second (GSa/s) with 1 GSa/s devices costing less than $100.

The first goal is to digitally sample the radio frequency (RF) signal as close to the antenna connection or focal point as possible. This reduces the total analog path length and amplifier gain required along with their associated temperature-dependant amplitude and phase fluctuations. Analog signal conditioning is still required to amplify the signal to the level required by the ADC and to limit the signal bandwidth to less than half of the digital sample rate to avoid sample aliasing.

The rms noise voltage at the input of a cryogenic receiver with a system temperature of 20 Kelvin and a bandwidth of 100 MHz is about 1 microvolt, whereas the ADC input signal level must be about 10 millivolts, so about 80 dB of net analog gain is required. Typical current receivers have 120 dB or more total gain to overcome multiple conversion, filter, and transmission line losses. A simplified receiver system will be more stable in at least two ways: less total gain and a much shorter signal path between the antenna terminals and the digitizer.

Direct digitization of the RF signal without analog frequency conversion is a possible strategy, particularly at low to modest frequencies, but frequency tuning flexibility, strong signal rejection, and low power consumption often makes analog conversion to a lower frequency band before digitization more attractive. It also allows the approximately 80 dB of gain to be divided between input and baseband frequency bands. Ensuring stability of 80 dB of gain at one frequency in a small space is a design problem one would prefer to avoid.

Most current centimeter-wave receivers employ several frequency conversion stages to allow wide tuning ranges with good image rejection, but each conversion stage adds spurious mixing products that can be quite troublesome. Limiting the analog portion of the receiver to one frequency conversion will produce a cleaner signal, and this is now feasible with digital signal processing to maintain excellent rejection of out-of-band signals. We briefly describe this strategy later in this white paper.

Immediate modulation of the digital data stream onto an optical fiber allows it to be transported away from the part of the antenna where space, power, and weight carrying capacity are at a premium. Optical fibers are small, light, and relatively insensitive to environmental conditions. Their bandwidth is limited only by the laser transmitters and diode receivers on either end.

Essentially all of the digital signal processing and data transport hardware components needed in radio astronomy applications is available from industry at continuously decreasing cost per unit bandwidth. However, many of the sophisticated packetizing, routing, and error-correction schemes needed for robust long-distance data transmission are not necessarily needed in the tens to thousands of meters distance from the antenna feed to the data processing components. The engineer's task, which is quite substantial, is to develop and implement algorithms that take advantage of the known statistical characteristics of radio astronomy signals to maximize the data throughput for a given volume, power, and weight, particularly on the transmit end of the fiber.

## Current State of the Art

The best-performing microwave and millimeter-wave heterodyne receivers in use today generally consist of large assemblies of discrete connectorized parts. A good example of one such receiver is the Q-Band 4-Beam Receiver on the Green Bank Telescope (Figure 1). Although providing unsurpassed performance, receivers such as these are clearly too bulky to be used in tightly-packed focal plane arrays, or possibly even by themselves on small dishes.

Possibly the most densely packed heterodyne receiver array constructed to date is the SEQUOIA system built at the University of Massachusetts for the Five College Radio Astronomy Observatory. This is a 16-pixel array with an input frequency range of 85-115 GHz and 16 analog output signals in the 5-20 GHz band. Two, single-polarization, 16-horn arrays with a wire grid and mirror were combined to receive both orthogonal polarizations. This array used MMIC amplifier technology, but most of the analog receiver components were located away from the telescope focus. This array receiver performed extremely well, but expansion to more than 16 pixels proved problematic[1].

More recently, a 60-pixel, dual-polarization horn array for 18-26 GHz has been proposed, and a 7-pixel prototype for this array is now under construction at the National Radio Astronomy Observatory. It uses connectorized components carefully arranged to fit behind the shadow of a compact corrugated feedhorn, thus permitting them to be used in a tightly packed focal plane array without resorting to any additional integration of components, as shown in Figure 2. Unfortunately, while the receiver dimensions have been greatly reduced laterally, they have also grown vertically. Size and mass therefore remain a critical issue for arrays such as this.

## Beyond the State of the Art

We believe the key to more effectively miniaturizing high-performance radio astronomy instruments is to transfer as many of the receiver functions as possible from the analog to digital signal domain. This is even true in the cryogenic front-end ahead of the first low-noise amplifier. For example, two independent, orthogonal polarizations from the sky are typically converted from degenerate modes in a dual-mode waveguide at one impedance to independent coaxial cable or rectangular waveguide modes at a different impedance. These mode and impedance transformations require minimum physical dimensions, set by fundamental physics. Wider bandwidths and tighter performance tolerances require even larger dimensions. By using four coaxial probes into the waveguide in a symmetric configuration, as shown in Figure 3a, and combining signals from opposing probes in digital arithmetic the size of the orthomode transducer (OMT) can be reduced and the isolation of the two polarizations can be greatly improved with calibrated signal combining coefficients. The penalty is the need for four low-noise amplifiers instead of two, but these are physically small. A more radical idea is to use only three probes and amplifiers, as shown in Figure 3b. The signal combing arithmetic is less intuitive, but it is essentially free. The analog equivalent of the three-probe OMT has never been tried, to our knowledge.

[1] http://www.astro.umass.edu/~fcrao/instrumentation/sequoia/seq.html

The waveguide device for converting from intrinsic linear to circular polarization (usually a phase-shifter preceding the OMT) can be eliminated entirely and its function done more accurately digitally. The final result is a dramatic reduction in the cold mass of a radio astronomy receiver. The short electrical length between the probes and the amplifiers may even eliminate the need for RF isolators. With precise calibration, polarization isolation in excess of 40 dB should be achievable, *with no compromise in front-end bandwidth or sensitivity*. High performance waveguide OMTs and polarizers typically provide isolations on the order of 20 to 25 dB. The reflector antenna to which the receiver is attached adds its own polarization signature, which could be included in the OMT signal processing, but this may remain as part of astronomical calibration and image processing algorithms. With better OMT and polarizer the image processing correction terms will be smaller.

Another area in which digital signal processing can help is frequency downconversion. Most heterodyne radio astronomy receivers incorporate at least two, often more, frequency conversion (mixing) stages to provide adequate frequency selectivity while tuning over broad bandwidths. Not only is this complex, but the use of multiple independent local oscillators (LO) opens the door for spurious mixing products to leak into the signal path. A better solution is to use a sideband-separating mixer to go from RF to a near-zero intermediate frequency (baseband IF) in one step. Once again, traditional analog components prove to be the limiting factor. A sideband-separating mixer requires a phase-quadrature power division of the LO signal and a broadband phase-quadrature combiner at IF. The net phase- and magnitude-imbalance of these power dividers and combiners typically limit sideband isolation to 20 dB or less, worse if the receiver must tune over a wide frequency range. This type of mixer is common at millimeter wavelengths, bit poor sideband separation has ruled out its use for in applications where radio frequency interference (RFI) is a problem.

The solution is to move the IF signal combiner of a sideband-separating mixer into the digital domain. Thus, the I and Q output signals of the mixer pair, shown in Figure 4, are digitized separately and then recombined using calibrated weighting coefficients. As with the OMT, the weighting coefficients can be optimized, not only to implement a mathematically perfect IF hybrid, but to compensate for phase and amplitude errors in the analog components. Again, the ultimate performance depends only on the resolution of the signal processing arithmetic and the stability of the analog hardware. We have demonstrated sideband isolation in excess of 50 dB in prototype mixers.

This approach does not require a larger number of mixers or increase the digital bit-rate into the backend for a given total IF bandwidth. It merely replaces two different mixers in a super-heterodyne scheme with two identical ones, and splits one analog to digital convertor (ADC) into two ADC's with half the sample rate. Although the number of digital channels has been doubled, the total digital signal bandwidth is exactly the same.

## Streamlining the Transition to Fiber

As has already been stated, the limiting performance of the digital-enhancement techniques described above depends in large part on the stability of the analog portion of the front-end. Component integration and elimination of connectors are well-proven strategies for ensuring that stability, in addition to the obvious size and weight advantages. It is our goal to integrate the conversions from analog to digital and from copper to fiber in a single compact package attached to the output of the cryogenic part of the front-end.

Doing so obviously requires overcoming some challenges. First, the samplers must be extremely well isolated from the analog components in order to avoid self-interference from high-speed digital signal transients. Traditional "conservative" techniques for doing this, by putting the digital portion in a physically-separated, shielded enclosure, are usually quite bulky and inefficient. However, this need not be the case. In principle, all that is required to isolate signals in two separate cavities is a metal partition which is only a few RF skin depths thick. There is no reason the analog and digital cavities could not be located very close to one another, and even cut out of the same metal housing.

To save space and power dissipation in the receiver package the digitized signal bits should be placed on an optical fiber with as little processing as is consistent with a robust data link. The thin fiber(s) will then transport the signals away from the antenna to a central processing facility that will house the bulky and power-hungry digital signal processing (DSP) hardware. Most of the burden of managing the communication across the link should be placed at the receiving end of the fiber, in the central DSP facility. This distinguishes our application from most commercial fiber optic solutions, where a great deal of encoding, formatting, and framing is done at the transmit end.

The very nature of astronomical data may be useful in simplifying the transmit side. For example, a general-purpose data link cannot make any assumptions about data content and must allow for cases where a long string of data bits could be all zeros or ones. All commonly deployed serial data links use the data itself to recover the data “clock” that marks the transitions between consecutive bits. If there are no data transitions for long periods, the recovered clock loses synchronization. To avoid this, the data are scrambled at the transmit end with a known algorithm to assure transitions and descrambled at the receive end with the reverse algorithm. However, in contrast to conventional data streams, digitally sampled astronomical receiver signals constitute nearly Gaussian-distributed noise. Clock/Data Recovery (CDR) circuits with Consecutive Identical Digit (CID) ratings of greater than 2000 bits are readily available, and the likelihood of an astronomical data stream comprising 2000 identical digits is infinitesimally small.

General purpose serial data links must also be able to detect message boundaries, work with any length of cable, and survive adaptive routing from computer to computer that may or may not preserve the original packet order. Radio astronomy data links are generally fixed and the data stream contiguous with slowly changing statistics. Data word boundaries can be detected at the receiving end by looking at bit statistics. The most significant bits will look very different from the least significant bits of the neighboring word. If two correlated data streams are mistakenly offset by one or more words, this will be seen as quantized phase and delay offsets in their correlation products. The only requirements are that these offsets be initially determined and that slow changes in cable lengths tracked in real time with signal processing on the receive end of the data links. The current state of the art for data links in radio astronomy is the system used in ALMA and EVLA. These use modified versions of the formatting and clock recovery techniques mentioned above for extremely robust link performance, and much is to be learned from that experience. However, we believe there is much to be gained with research into low-overhead, point-to-point digital data transfer strategies.

## Final Technology Goals and Development Milestones

The various enhancements and simplifications described above go hand-in-hand toward realizing a well-optimized modern receiver architecture that is compact, inexpensive, reliable, and mass producible without comprising performance or versatility in any significant way. The following list of milestones is a step-by-step development program toward achieving the final goals.

1. Digital Sideband Separating Mixer (DSSM) -- This initial proof of concept has been achieved in the form of an L-Band (1200-1700 MHz) downconverter with 500 MHz total IF Bandwidth, the digital recombination performed in software as post-processing. The results are shown in Figure 5.
2. DSSM With Integrated ADCs -- This milestone will demonstrate the compact integration of high-gain analog and high-speed digital components in a common housing with sufficient isolation to avoid self-interference.
3. Digital Ortho-Mode Transducer (DOMT) -- This will be the initial proof of concept for digital polarization isolation. Prototypes will be developed at X-Band using both the three- and four-probe configurations. As with the above milestones, digital signal recombination will be performed in post-processing at this stage.
4. Minimal-Overhead Radio Astronomy Photonic Link -- This will demonstrate wide bandwidth transport of data over an optical fiber with a minimum of overhead for formatting and packet-framing functions at the transmit end. A very simple block diagram of what the link may look like is shown in Figure 6.
5. RF-Input/Fiber-Output Warm Receiver -- This will involve marrying the above concepts into a single module that encapsulates the conversion from RF to baseband, from analog to digital, and from copper to fiber within a shared compact housing.
6. Real-Time Signal Recombination -- The signal recombination arithmetic will be programmed into an FPGA, allowing for the first time the digital sideband separation and polarization isolation to be performed not in post-processing, but in real-time.
7. Complete Next-Generation Receiver at X-Band -- A complete X-Band receiver will be tested on the Green Bank Telescope (GBT), including a cryogenic front-end module with either a three-probe or four-probe OMT, an RF-input/fiber-output warm receiver module, a minimal-overhead photonic link, and real-time signal recombination hardware feeding into the backend spectrometer.

8. Next-Generation FPA at W-Band -- Two or more complete next-generation receivers will be constructed as above, this time at W-Band, and tested as a small prototype focal plane-array with approximately 2.3cm spacing.
9. Beam-Forming Array Prototype at X-Band -- The X-Band next-generation receivers will be replicated, with a modified DOMT configured to mate with crossed-dipoles instead of a feed horn, and tested as a beam-forming array with approximately 2cm spacing. In addition to proving an unprecedented level of miniaturization at X-Band, it will also test relative receiver chain stability between channels.
10. Ultra-Wideband Next-Generation Receivers -- As higher-speed ADCs, FPGAs, and Optoelectronic components become available, subsequent generations of the receivers described in this white paper will be constructed with increasing instantaneous IF bandwidths.

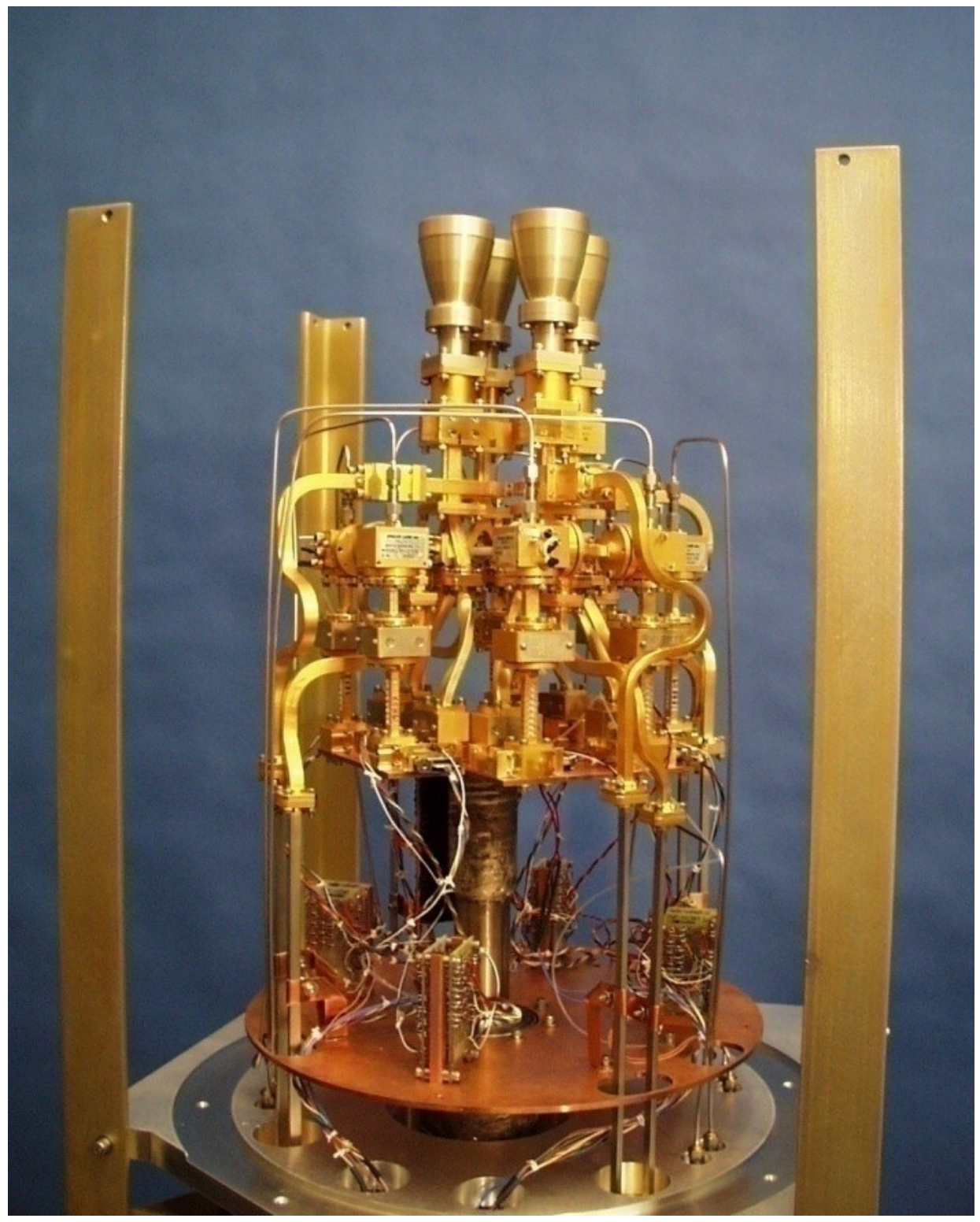

FIGURE 1. Q-BAND RECEIVER ON THE GREEN BANK TELESCOPE.

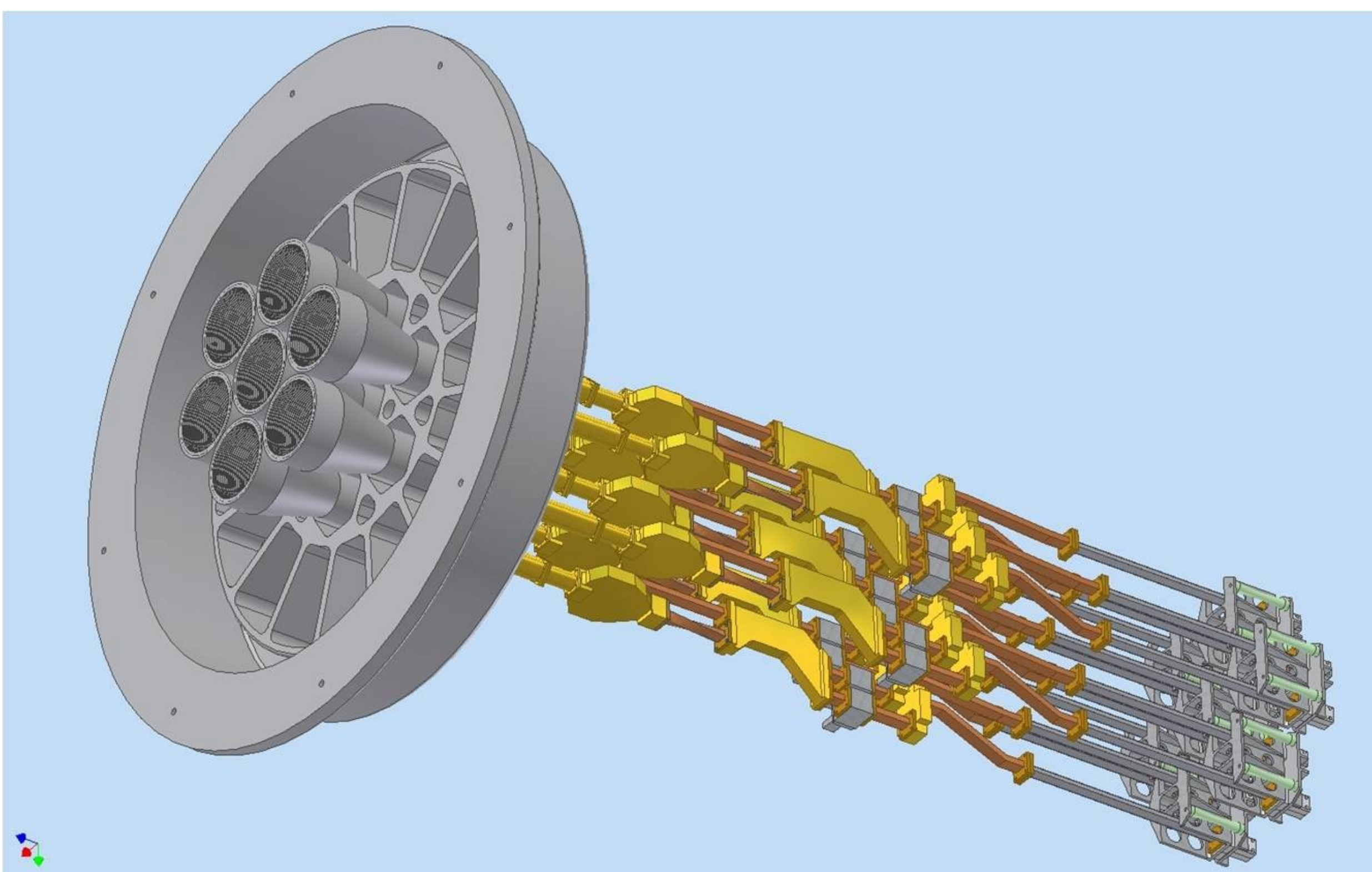

FIGURE 2. DRAWING OF THE SEVEN BEAM K-BAND FOCAL PLANE ARRAY FOR THE GBT.

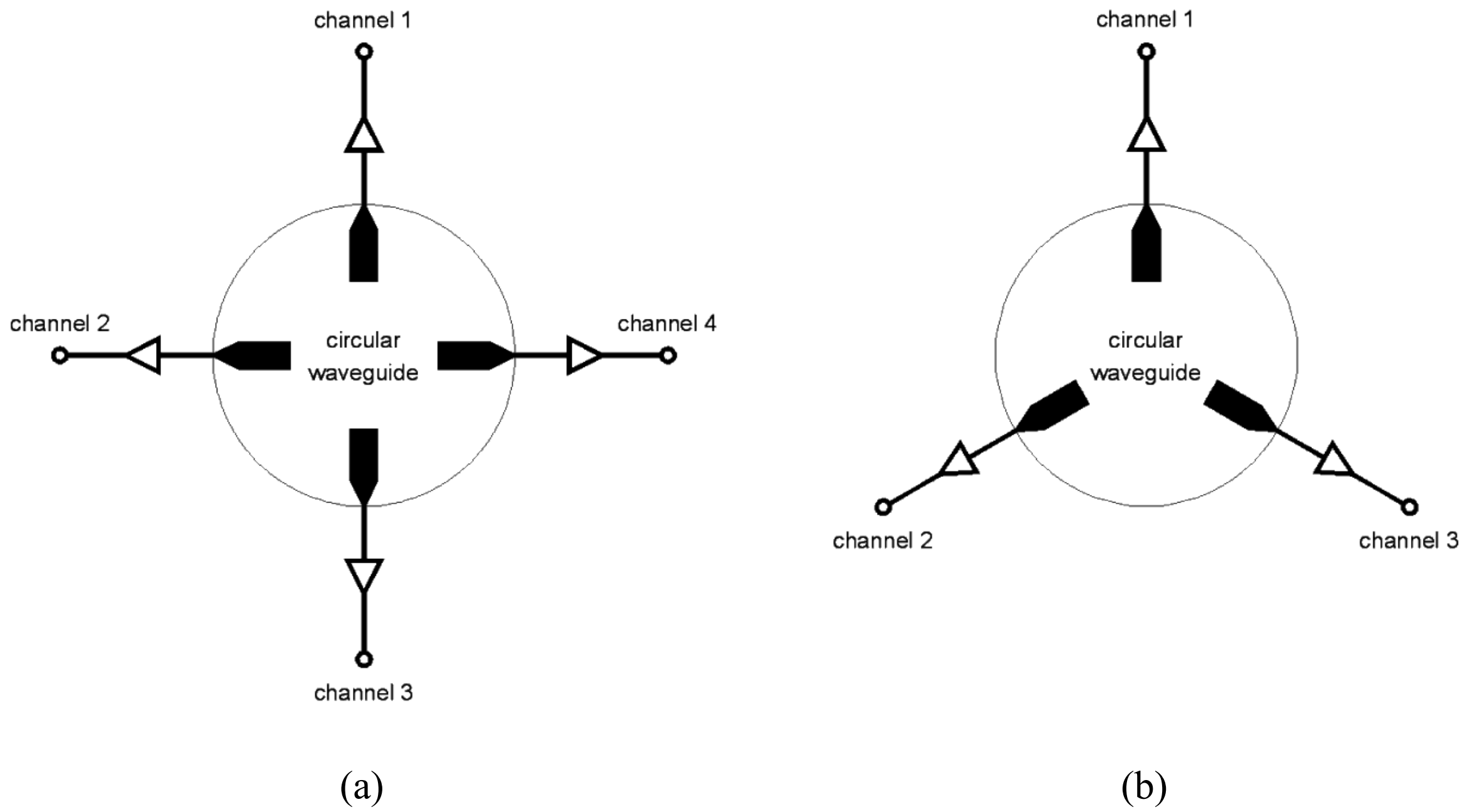

FIGURE 3.ANALOG PORTION OF A DIGITALLY-ENHANCED OMT. THE OUTPUTS ARE PROCESSED INDEPENTLY AND THEN MATHEMTICALLY RECOMBINED IN THE DIGITAL BACKEND.

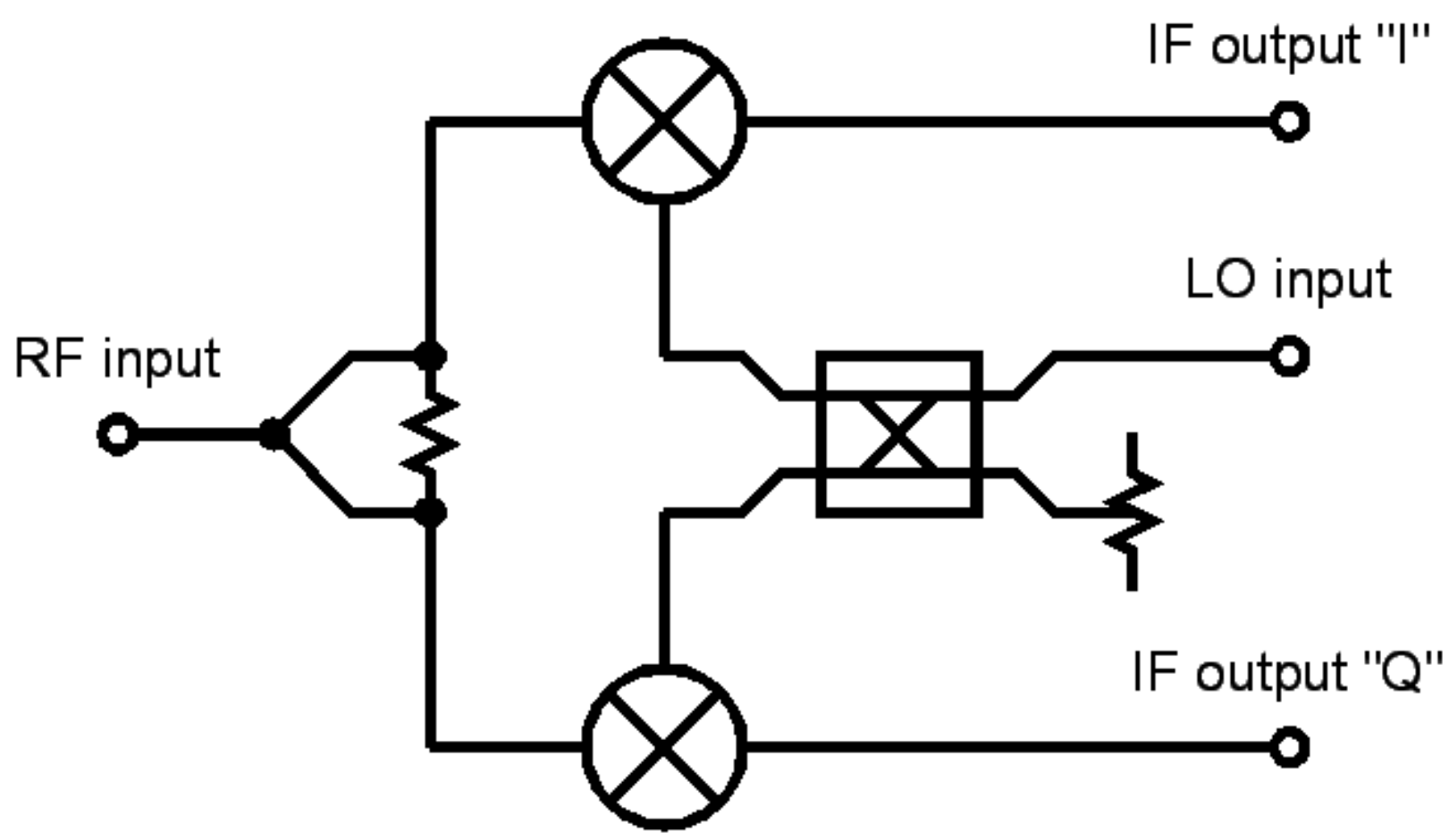

FIGURE 4. ANALOG PORTION OF A DIGITALLY-ENHANCED SIDEBAND-SEPARATING MIXER. THE OUTPUTS ARE PROCESSED INDEPENDENTLY AND THEN MATHEMATICALLY RECOMBINED IN THE DIGITAL BACKEND.

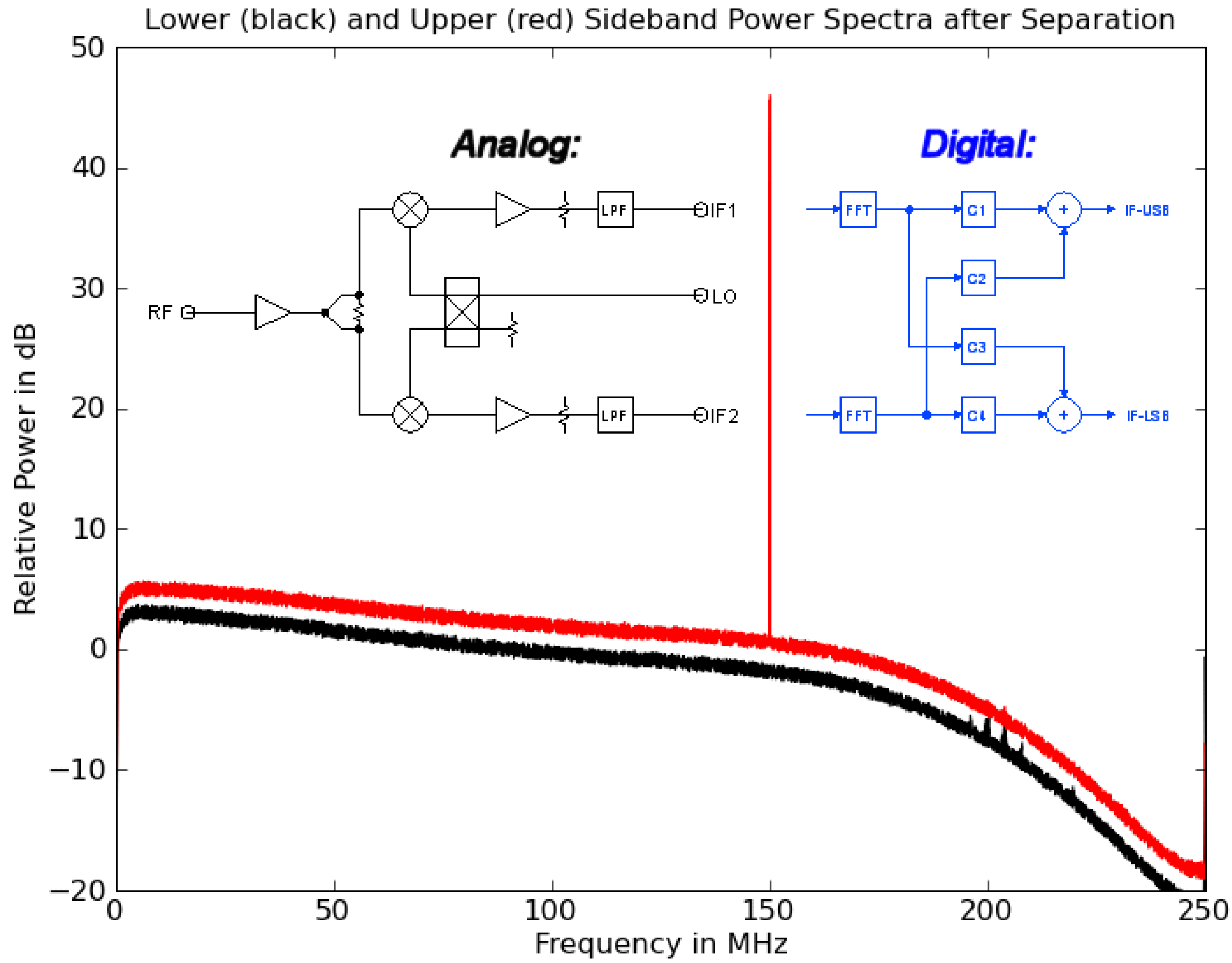

FIGURE 5. TYPICAL SIDEBAND-SEPARATION MEASUREMENT USING L-BAND PROTOTYPE. GAIN OFFSET ADDED FOR CLARITY.

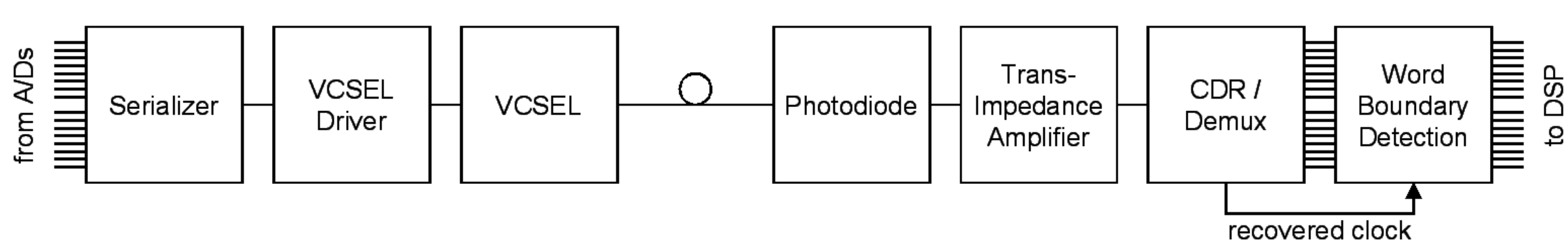

FIGURE 6. DIAGRAM OF A MINIMAL PHOTONIC LINK FOR RADIO ASTRONOMY RECEIVERS.